\documentclass{aa}  
\usepackage{graphicx}
\usepackage{txfonts}
\usepackage{hyperref}
\begin{document} 

\title{Beneath the surface: $>$85\% of z$>$5.9 quasi-stellar objects in massive host galaxies are UV-faint}
\author{Rychard J. Bouwens\inst{1}
\and Eduardo Ba{\~n}ados\inst{2}
\and Roberto Decarli\inst{3}
\and Joseph Hennawi\inst{1,4}
\and Daming Yang\inst{1}
\and Hiddo Algera\inst{5,6,7}
\and Manuel Aravena\inst{8}
\and Emanuele Farina\inst{9}
\and Anniek Gloudemans\inst{9}
\and Jacqueline Hodge\inst{1}
\and Hanae Inami\inst{6}
\and Jorryt Matthee\inst{10}
\and Romain Meyer\inst{11}
\and Rohan P. Naidu\inst{12}
\and Pascal Oesch\inst{11,13,14}
\and Huub J.A. R{\"o}ttgering\inst{1}
\and Sander Schouws\inst{1}
\and Renske Smit\inst{15}
\and Mauro Stefanon\inst{16,17}
\and Paul van der Werf\inst{1}
\and Bram Venemans\inst{1}
\and Fabian Walter\inst{2}
\and Yoshinobu Fudamoto\inst{18,19}}
\institute{$^1$ Leiden Observatory, Einsteinweg 55, NL-2333 CC Leiden, The Netherlands, \email{bouwens@strw.leidenuniv.nl}\\
  $^2$ Max-Planck-Institut f{\"u}r Astronomie, K{\"o}nigstuhl 17, 69117 Heidelberg, Germany,\\
  $^3$ INAF–Osservatorio di Astrofisica e Scienza dello Spazio, via Gobetti 93/3, I-40129, Bologna, Italy\\
  $^4$ Department of Physics, Broida Hall, University of California, Santa Barbara, Santa Barbara, CA 93106-9530, USA\\
  $^5$ Institute of Astronomy and Astrophysics, Academia Sinica, 11F of Astronomy-Mathematics Building, No.1, Sec. 4, Roosevelt Rd, Taipei 106216, Taiwan, R.O.C.\\
  $^6$ Hiroshima Astrophysical Science Center, Hiroshima University, 1-3-1 Kagamiyama, Higashi-Hiroshima, Hiroshima 739-8526, Japan\\
  $^7$ National Astronomical Observatory of Japan, 2-21-1, Osawa, Mitaka, Tokyo, Japan\\
  $^8$ Instituto de Estudios Astrof{\'\i}sicos, Facultad de Ingenier{\'\i}a y Ciencias, Universidad Diego Portales, Av. Ej{\'e}rcito 441, Santiago, Chile\\
  $^9$ International Gemini Observatory/NSF NOIRLab, 670 N A’ohoku Place Hilo, HI 96720, USA\\
  $^{10}$ Institute of Science and Technology Austria (ISTA), Am Campus 1, 3400 Klosterneuburg, Austria\\
  $^{11}$ Department of Astronomy, University of Geneva, Chemin Pegasi 51, 1290 Versoix, Switzerland\\
  $^{12}$ MIT Kavli Institute for Astrophysics and Space Research, 70 Vassar Street, Cambridge, MA 02139, USA\\
  $^{13}$ Cosmic Dawn Center (DAWN), Copenhagen, Denmark\\
  $^{14}$ Niels Bohr Institute, University of Copenhagen, Jagtvej 128, København N, DK-2200, Denmark\\
  $^{15}$ Astrophysics Research Institute, Liverpool John Moores University, 146 Brownlow Hill, Liverpool L3 5RF, UK\\
  $^{16}$ Departament d’Astronomia i Astrof{\'\i}sica, Universitat de Val{\`e}ncia, C. Dr. Moliner 50, E-46100 Burjassot, Val{\`e}ncia, Spain\\
  $^{17}$ Unidad Asociada CSIC ”Grupo de Astrof{\'\i}sica Extragal{\'a}ctica y Cosmolog{\'\i}a” (Instituto de Física de Cantabria - Universitat de Val{\`e}ncia), Spain\\
  $^{18}$ Steward Observatory, University of Arizona, 933 N Cherry Avenue, Tucson, AZ 85721, USA\\
  $^{19}$ Center for Frontier Science, Chiba University, 1-33 Yayoi-cho, Inage-ku, Chiba 263-8522, Japan}

\date{Received June 2025; Accepted soon after} \authorrunning{Bouwens
  et al.}  \titlerunning{Beneath the surface: $>$85\% of z$>$5.9
  quasi-stellar objects in massive host galaxies are UV-faint}
\abstract { We used [C II]$_{158\mu\textrm{m}}$ observations of a
  large quasi-stellar object (QSO) sample to segregate sources by host
  galaxy mass to identify those in the most massive hosts. The [CII]
  luminosity is a known tracer of molecular gas. It was taken as a
  proxy for the host mass and was used to rank 190 QSOs at $z>5.9$,
  spanning a UV luminosity range of 6 mag
  ($-28<M_{UV,AB}<-22$). Atacama Large Millimeter Array (ALMA) data
  from a cycle-10 program were particularly valuable. They provide
  [CII] coverage for 46 UV-faint ($M_{UV,AB}>-24.5$) and 25 especially
  UV-faint ($M_{UV,AB}>-23.5$) QSOs. This improved the statistics by
  five and six times, respectively.  Taking massive host galaxies to
  be those where $L_{\text{[CII]}}$ $>1.8\times10^9$ L$_\odot$ (median
  $L_{[CII]}$ of UV-bright QSOs), we identified 63 QSOs, including 13
  that are UV-faint and 7 that are especially UV-faint.  Using these
  selections and recent QSO luminosity functions (LFs), we present the
  first characterization of UV luminosity distribution for QSOs in
  massive host galaxies and quantify [CII] LFs for the UV-bright and
  UV-faint QSOs.  While $\sim$3\% of the massive-host QSOs are
  UV-bright ($M_{UV,AB}<-26$), $\gtrsim$85\% are UV-faint
  ($M_{UV,AB}>-24.5$).  This wide dispersion in UV luminosities
  reflects variations in dust obscuration, accretion efficiency, and
  black hole mass. Even though spectroscopy is needed for definitive
  conclusions, the black hole mass appears to be the dominant factor
  driving variations in the UV luminosity, based on 34 [CII]-luminous
  ($L_{\text{[CII]}}>1.8\times10^9$ L$_{\odot}$) QSOs distributed
  across a baseline of $\sim$3 mag in UV luminosity and with measured
  $M_{BH}$.  At $M_{UV,AB}\sim-23$, the median extrapolated $\log_{10}
  (M_{BH}/M_{\odot})$ is $8.1\pm0.4$, consistent with the local
  relation.  Supermassive black holes (SMBHs) in UV-bright QSOs thus
  appear to be $\sim$15$_{-9}^{+25}$$\times$ more massive than is
  typical for massive host galaxies at z$\sim$6.}

   \keywords{Galaxies: active, (Galaxies:) quasars: emission lines, Galaxies: high-redshift, Galaxies: star formation}

   \maketitle
   
\section{Introduction}

The discovery of supermassive black holes (SMBHs; 10$^9$ M$_{\odot}$)
in bright quasi-stellar objects (QSOs) just 700-900 Myr after the Big
Bang has been an exciting frontier in extragalactic cosmology, as have
the questions they inspire regarding the formation of such SMBHs,
which apparently require very massive ($>$10$^4$ M$_{\odot}$) black
hole (BH) seeds or super-Eddington accretion
\citep[e.g.,][]{Fan2001,Haiman2001,Volonteri2003,Mortlock2011,Banados2018,Volonteri2021,Fan2023,Wang2024}.
Results from {\it James Webb Space Telescope} ({\it JWST}) data
suggest the formation of massive BHs in the early Universe is
ubiquitous
\citep[e.g.,][]{Kocevski2023,Harikane2023_AGN,Larson2023,Maiolino2024,Matthee2024_LRDs,Labbe2023_LRDs,Kokorev2024,Akins2024}
and occur at redshifts as high as z = 10.2 \citep{goulding2023}.

The discovery of a widespread population of SMBHs in the early Universe has inspired the development of models to explain their buildup.  One challenge in these efforts is to account for the large variability in the detectability of individual SMBHs, which is driven by factors such as episodic gas accretion and obscuration by dust \citep[e.g.,][]{Urry1995,Padovani2017}. To mitigate this, it is useful to link the SMBH growth to that of their host galaxies, leveraging our general understanding of how stellar and halo masses evolve.  Unlike SMBHs, host galaxies are consistently observable in surveys and tend to build up their halo, stellar, and gas content in a generally predictable fashion.  By measuring host galaxy masses across cosmic time and rank ordering the galaxies according to mass \citep[e.g.,][]{Vale2004,Conroy2006}, we can connect SMBHswith galaxies of a given mass, providing us with a framework for quantifying the growth of SMBHs across cosmic time.

The entire enterprise of characterizing host galaxy properties for bright quasi-stellar objects (QSOs) to facilitate a QSO-host galaxy connection is challenging because it is hard to distinguish the host galaxy light from that of the QSO itself \citep[e.g.,][]{Schramm2008,Matsuoka2014,Ciesla2015}.  Fortunately, enormous progress is now being made with JWST through its exceptional sensitivity and resolving power, which provides a robust separation of QSO and host galaxy light, and thus, more reliable stellar mass determinations \citep[e.g.,][]{Ding2023}.  Despite these advances, it is still difficult to characterize the host galaxy properties for the brightest QSOs \citep{Yue2024}, even with JWST.  

An appealing alternative to using rest-frame optical light to probe host galaxy properties is far-IR line emission from tracers such as [CII]$_{158\mu\text{m}}$. The [CII] line is a well-established tracer of the molecular gas mass \citep{Zanella2018, Madden2020, Vizgan2022_HII}, neutral gas mass \citep{Heintz2021, Vizgan2022_HI}, and star formation rates \citep[SFRs;][]{deLooze2014}. It also offers several observational advantages: it is readily detectable with facilities such as Atacama Large Millimeter Array (ALMA) \citep[e.g.,][]{Decarli2018,Mazzucchelli2025} and Northern Extended Millimetre Array (NOEMA) \citep[e.g.,][]{Khusanova2022,Banados2024}, is largely unaffected by dust obscuration, free from contamination by bright QSO emission at the same frequency, and minimally affected by the QSO in powering emission within photodissociation regions \citep[e.g.,][]{Stacey2010, Sargsyan2014}.  Given the expected correlation between [CII] luminosity and host galaxy molecular gas mass \citep{Zanella2018,Vizgan2022_HII}, this line provides a powerful tool for distinguishing QSOs according to the mass of their host galaxies (see, however, \citealt{Kaasinen2024}), and this allows us a detailed study of QSO demographics in massive systems.

To map out the demographics of QSOs in massive host galaxies using far-IR diagnostics such as [CII], it is useful to consider QSOs from a range in UV luminosity that is as wide as possible. We therefore drew from traditional UV-bright type 1 QSO samples and also from fainter selections of QSOs in the rest-UV such as are pursued by SHELLQs \citep[Subaru High-z Exploration of Low-Luminosity Quasars;][]{Matsuoka2016_SHELLQSI}.   This ensured that type 1 QSOs with a range of BH masses, Eddington ratios, and dust obscuration were included\footnote{This only samples moderately obscured QSOs to the extent they are included in type 1 QSO selections.} and provided a much more representative selection of SMBHs in massive galaxies than the UV-bright population provides. This topic that has been extensively discussed in the literature \citep[e.g.,][]{Willott2005,Lauer2007,Reines2015,Shankar2020,Wu2022,Li2022}.  Ideally, the selections used for these purposes should even include the so-called little red dot population (\citealt{Matthee2024_LRDs,Greene2024})  that was identified with JWST \citep[e.g.,][]{Kocevski2023, Labbe2023_LRDs}, but their consistent faintness in the far-IR and [CII]$_{158\mu\text{m}}$ \citep[e.g.,][]{Akins2024,Setton2025,Xiao2025,Casey2025} suggests that they are associated with much lower-mass host galaxies than the galaxies hosting UV-bright QSOs.

\begin{figure}
   \includegraphics[width=9.1cm]{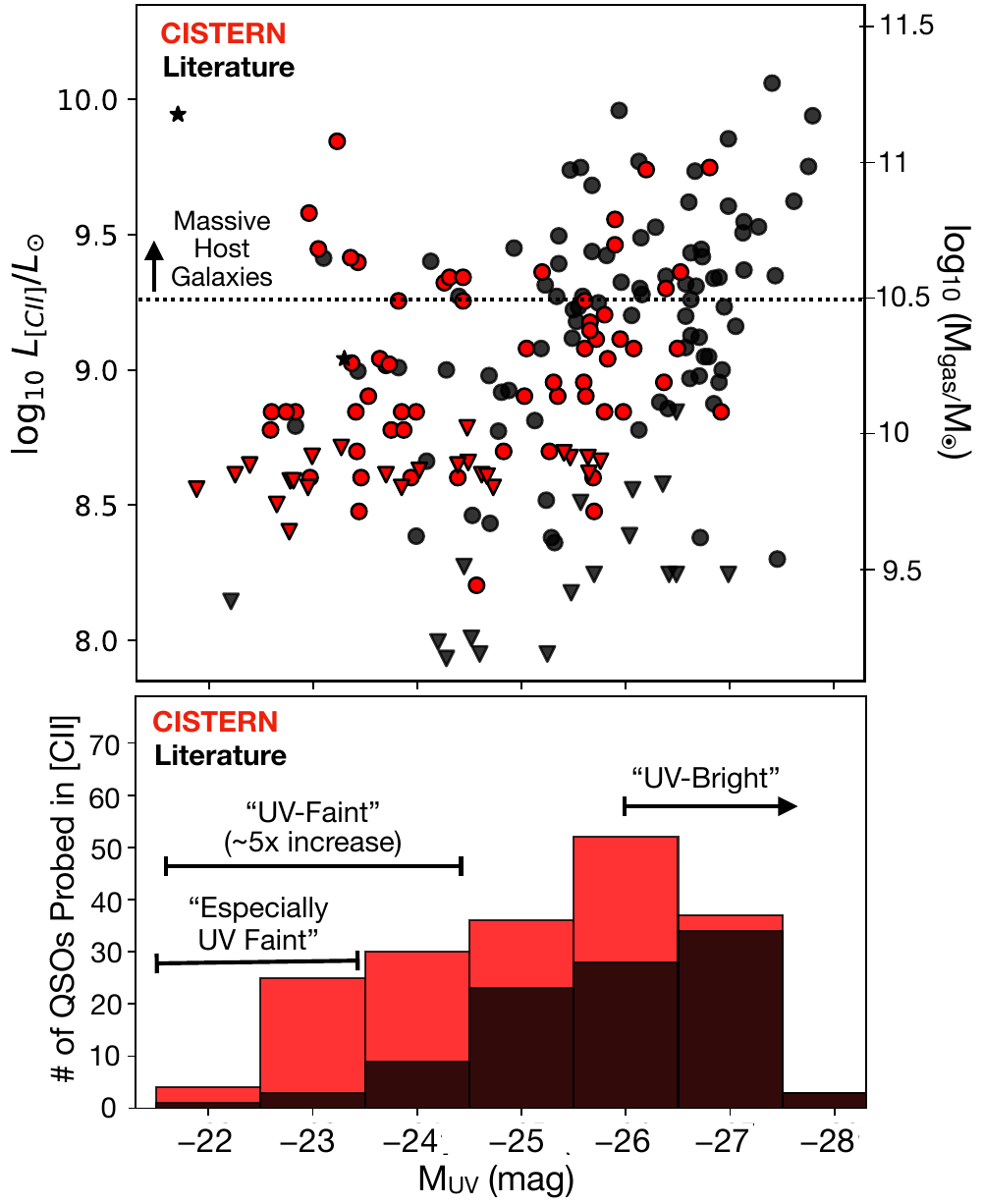}
   \caption{\textit{Top}: Measured [CII] luminosities
     versus $M_{UV}$ luminosities for $z>5.9$ QSOs from CISTERN
     (red circles) and from the literature (black circles).  The solid downward-pointing triangles indicate
     $5\sigma$ upper limits on the [CII] luminosities of QSO where no
     line is detected with ALMA.  The black stars correspond to the
     \citet{Fujimoto2022_redQSO} and \citet{Endsley2023_CII} QSOs that
     were identified in deep multiwavelength data over legacy fields
     and not from wide-area QSO searches.  We show along the right
     vertical axis the approximate gas mass that is expected for
     the host galaxy of a given [CII] luminosity using the canonical
     \citet{Zanella2018} relation. About 21\%\ of the UV-faint QSOs
     show luminous [CII] emission and thus appear to reside in massive
     host galaxies. \textit{Bottom}: Number of [CII]
     observations for $z>5.9$ QSOs versus UV luminosity from the
     literature (filled black histogram) and from CISTERN
     (filled red histogram). The observations from
     CISTERN increased the
     coverage of [CII] in UV-faint ($M_{UV,AB}>-24.5$) QSOs with ALMA by about five times.
   \label{fig:muvcii}}
    \end{figure}

\begin{table}
\setlength{\tabcolsep}{5.5pt}
\caption{Observational datasets used to probe [CII] emission from z$>$5.9 QSOs.\vspace{-0.3cm}}
\label{tab:datasets}      
\centering          
\begin{tabular}{c c c c c}      
\hline\hline     
     &          & \multicolumn{3}{c}{\# QSOs} \\
Data & Redshift & \multicolumn{3}{c}{UV Luminosity$^a$}\\
Set &  Range &  Faint & Intrmd & Bright \\
\hline
\multicolumn{5}{c}{This Work}\\
CISTERN & 5.9-7.0 & 46 & 32 & 8\\
2019.1.00074.S & 6.1-6.5 & 4 & 6 & 0 \\\\
\multicolumn{5}{c}{Earlier Work}\\
\citet{Willott2017_faintQSO} & 6.0-6.2 & 3 & 0 & 0 \\
\citet{Izumi2018} & 5.9-6.4 & 3 & 0 & 0 \\
\citet{Decarli2018} & 6.0-6.4 & 0 & 2 & 10 \\
D18 Lit$^b$ & 6.0-6.7 & 0 & 4 & 4 \\
\citet{Izumi2019} & 6.1-6.1 & 1 & 2 & 0 \\
\citet{Eilers2020} & 5.9-6.5 & 0 & 2 & 4 \\
\citet{Venemans2020} & 6.0-7.5 & 0 & 5 & 20 \\
\citet{Izumi2021} & 7.0-7.1 & 1 & 0 & 0 \\
\citet{Izumi2021_redQSO} & 6.7-6.8 & 1 & 0 & 0 \\
\citet{Fujimoto2022_redQSO} & 7.1-7.2 & 1 & 0 & 0 \\
\citet{Endsley2023_CII} & 6.8-6.9 & 1 & 0 & 0 \\
\citet{Wang2024}$^c$ & 6.5-7.6 & 0 & 14 & 16 \\
\hline
Total & 5.9-7.6 & 61 & 67 & 62 \\
\hline
\end{tabular}
\tablefoottext{a}{Faint, intrmd, and bright subsamples refer to QSOs with UV luminosities satisfying the criteria $-24.5\leq M_{UV,AB}$, $-26.0\leq M_{UV,AB}<-24.5$, and $M_{UV,AB}<-26.0$, respectively.}

\tablefoottext{b}{Literature compilation presented in Table 2 of \citet{Decarli2018}, including results from \citet{Walter2009}, \citet{Wang2013}, \citet{Willott2013}, \citet{Willott2015}, and \citet{Mazzucchelli2017}.}

\tablefoottext{c}{A magnification factor of 50 was adopted to correct the UV luminosity of J043947.08+163415.7 for gravitational lensing \citep{Fan2019}.}

\end{table}

The large selection of lower-luminosity QSOs recently obtained from
the CISTERN ([CII]-SelecTed Early universe RefereNce, PI: Bouwens)
cycle-10 ALMA program (Bouwens et al.\ 2026, in prep) has proven
particularly useful for obtaining [CII] coverage of z$\gtrsim$6 QSOs
over a wide range in UV luminosity.  CISTERN obtained ALMA coverage
probing the [CII] line for 46 UV-faint ($-24.5<M_{UV,AB}$) and 25
especially UV-faint ($-23.5<M_{UV,AB}$) QSOs at z$>$5.9, which increased
the [CII] coverage of z$\gtrsim$6 QSOs in these luminosity ranges by
five and six times, respectively. For QSOs of all
UV luminosities, [CII] coverage is available for 190 z$>$5.9 QSOs,
45\% of which are provided by the CISTERN program.  More information
on the CISTERN program and the basic results will be provided in
Bouwens et al.\ (2026, in prep).

With [CII] luminosity information, we segregated the 190 QSOs
in our CISTERN+archive sample according to the apparent mass of their
host galaxies.  Most of our attention was devoted to QSOs
associated with the most massive galaxies because they are ubiquitous
among UV-bright QSOs and are fairly frequent among even fainter
QSOs, as we show here (and see \citealt{Izumi2021}). In addition to the luminosity of the [CII] line, the mass of the host galaxies is also commonly probed using the dynamical mass (as inferred from
the width of [CII]) and using the far-IR luminosity
\citep[e.g.,][]{Willott2013,Izumi2019,Izumi2021,Neeleman2021}, and we considered these alternate measures in the massive host galaxy
samples we constructed.

We then leveraged published UV luminosity functions (LFs) for QSOs \citep[e.g.,][]{Matsuoka2018_QSOLF,Schindler2023,Matsuoka2023} to provide volume density information of the QSOs with [CII] coverage, allowing us to compute [CII] LF results for QSOs, and thus, to compare the prevalence of UV-bright and UV-faint QSOs in host galaxies of a given mass.  Additionally, the UV LF information allowed us to derive various QSO distribution functions for host galaxies of a given mass, for instance, including for the UV and bolometric luminosities. It is also possible to determine the BH mass and Eddington ratio distributions for QSOs in massive host galaxies, but this also requires spectroscopy for the sources that are used to construct the distribution functions. Because we lack such information for the fainter QSOs in our massive host galaxy sample, we explored an extrapolation of results from brighter QSOs to make a first estimate.

We provide a brief plan for the paper.  In Sect. 2 we present the observational data we used, and in Sect. 3 we present the distribution of [CII]$_{158\mu \text{m}}$ luminosities we measure versus UV luminosity for our composite CISTERN+archival sample and use this to map out the UV luminosity distribution for QSOs in massive host galaxies, while deriving [CII]$_{158\mu\text{m}}$ LF results for UV-bright and UV-faint QSOs.  Finally, making use of trends in $M_{BH}$ and $\lambda_{Edd}$ in UV luminosity, we present our derived $M_{BH}$ and $\lambda_{Edd}$ distributions for QSOs in massive host galaxies.  For simplicity and to facilitate comparison with other studies, we adopted the standard so-called concordance cosmology, with $\Omega_m = 0.3$, $\Omega_{\Lambda} = 0.7$, and $H_0 = 70$ km/s/Mpc.  All magnitude measurements are given using the AB magnitude system \citep{OkeGunn1983} unless otherwise specified.

\begin{table*}
\caption{Compilation of UV-faint QSOs in massive host galaxies$^a$ at z$>$5.9\vspace{-0.2cm}.}
\label{tab:massive_uv_faint}      
\centering          
\begin{tabular}{c c c c c c c c c}      
\hline\hline     
Source  &  &    & $M_{UV,AB}$ &  & $L_{\text{[CII]}}$ & FWHM$_{[\text{CII}]}$ & L$_{\text{IR}}$ & \\
 ID & R.A. & Decl. & (mag) & $z_{[\text{CII}]}$ & $/10^{9}$ $L_{\odot}$ & (km/s) & (10$^{12}$ L$_{\odot}$) & Ref$^b$\\
\hline\\
\multicolumn{9}{c}{Based on [CII]$_{158\mu\text{m}}$ Data From This Work (CISTERN)}\\
J132308.18+012619.2 & 13:23:08.18 & $+$01:26:19.2 & $-23.0$ & 6.026 & 3.8$\pm$0.1 & 283$\pm$9 & 3.8$\pm$0.1 & This Work, [1]\\
J144045.91+001912.9$^{c}$ & 14:40:45.91 & $+$00:19:12.9 & $-$23.2 & 6.549 & 7.0$\pm$0.1 & 389$\pm$10 & 8.4$\pm$0.3 & This Work, [1]\\
HSCJ142903$-$010443 & 14:29:03.08 & $-$01:04:43.4 & $-$23.3 & 6.796 & 2.8$\pm$0.2 & 355$\pm$18 & 2.5$\pm$0.1 & This Work, [2]\\
J132700.35+014147.7 & 13:27:00.35 & $+$01:41:47.7 & $-23.4$ & 6.217 & 2.6$\pm$0.2 & 166$\pm$7 & 5.2$\pm$0.2 & This Work, [1]\\
HSCJ114632$-$015438$^{d}$ & 11:46:32.66 & $-$01:54:38.3 & $-$23.5 & 6.158 & 2.5$\pm$0.1 & 135$\pm$6 & 1.7$\pm$0.1 & This Work, [3,4] \\
J145520.26+031833.0 & 14:55:20.26 & $+$03:18:33.0 & $-23.8$ & 6.053 & 1.8$\pm$0.1 & 372$\pm$27 & 1.9$\pm$0.1 & This Work, [1]\\
J122331.91+025721.5 & 12:23:31.91 & $+$02:57:21.5 & $-24.3$ & 6.258 & 2.1$\pm$0.3 & 547$\pm$89 & 4.0$\pm$0.2 & This Work, [1]\\
ILTJ2336+1842 & 23:36:24.69 & +18:42:48.7 & $-$24.3 & 6.578 & 2.2$\pm$0.1 & 297$\pm$17 & 3.1$\pm$0.2 & This Work, [5] \\
HSCJ225520+050343 & 22:55:20.78 & +05:03:43.3 & $-24.5$ & 6.172 & 1.8$\pm$0.2 & 257$\pm$24 & 5.1$\pm$0.2 & This Work, [6]\\\\
\multicolumn{9}{c}{From the Literature}\\
J0958+0139$^{d}$ & 09:58:58.27 & +01:39:20.2 & $-$21.7 & 6.853 & 8.8$\pm$0.8 & 652$\pm$17 & 9.1$\pm$0.5$^e$ & [7,8] \\
VIMOS2911  & 22:19:17.23 & +01:02:48.9 & $-$23.1 & 6.149 & 2.6$\pm$0.1 & 264$\pm$15 & 2.2$\pm$0.1$^e$ & [9] \\
J1243+0100 & 12:43:53.93 & +01:00:38.5 & $-$24.1 & 7.075 & 2.5$\pm$0.1 & 280$\pm$12 & 5.4$\pm$0.1$^e$ & [10] \\
J1205$-$0000$^{d}$ & 12:05:05.09 & $-$00:00:27.9 & $-$24.4 & 6.723 & 1.9$\pm$0.1 & 536$\pm$26 & 3.8$\pm$0.1$^e$ & [4,11] \\
\hline
\end{tabular}
\tablefoottext{a}{As defined by $L_{[\text{CII}]}>1.8\times10^{9}$ $L_{\odot}$}
\tablefoottext{b}{References: [1] = \citet{Matsuoka2022}, [2] = \citet{Matsuoka2018a}, [3] = \citet{Matsuoka2018b}, [4] = \citet{Kato2020}, [5] = \citet{Gloudemans2022}, [6] = \citet{Matsuoka2019a}, [7] = \citet{Endsley2022_CandidateAGN}, [8] = \citet{Endsley2023_CII}, [9] = \citet{Willott2017_faintQSO}, [10] = \citet{Izumi2021}, [11] = \citet{Izumi2021_redQSO}}
\tablefoottext{c}{Red near-IR colors of this source in ground-based near-IR data + WISE \citep{Matsuoka2022} suggest that this QSO is dust reddened (Bouwens et al.\ 2026, in prep).}
\tablefoottext{d}{Among the UV-faint QSOs that were suggested to be obscured by dust \citep{Kato2020,Endsley2022_CandidateAGN}.}
\tablefoottext{e}{To ensure consistency with the other QSOs in this study, the IR luminosities tabulated here were recomputed using the reported continuum fluxes for these QSOs and the \citet{Venemans2020} method.}
\end{table*}

\section{Observational data}

We made use of ALMA observations targeting the [CII] line from
190 QSOs at z$>$5.9.  Observations from the cycle-10 CISTERN
program (2023.1.00443.S: R.J. Bouwens et al. 2026, in prep)
provided us with key new information for this analysis, but we supplemented these observations with [CII] results on 104 QSOs from archival programs. The UV luminosities were taken for QSOs from \citet{Fan2023} where available and were otherwise computed based on the available photometry in the near-IR band closest to rest-frame 1450\AA$\,$(but which is not affected by absorption blueward of 1216\AA).  Table~\ref{tab:datasets} provides a convenient
summary of the datasets and QSOs included in this analysis.

New observations from the cycle-10 CISTERN program were key to this analysis because of the huge increase in the coverage of [CII] for QSOs that are relative UV-faint ($M_{UV,AB}>-24.5$).  Of the 87 $z\geq5.9$ QSOs with delivered data from the CISTERN program, 46 corresponded to lower-luminosity (or UV-faint) $M_{UV,AB}\geq -24.5$ QSOs, 32 corresponded to intermediate-luminosity $-26.0\leq M_{UV,AB}<-24.5$ QSOs, and 8 corresponded to high-luminosity $M_{UV,AB}<-26.0$ QSOs.  By contrast, 
published measurements on [CII]$_{158\mu\text{m}}$ only exist for 11 fainter QSOs ($M_{UV,AB}\geq -24.5$) from earlier
programs, only 9 of which were identified based on wide-area search
efforts.  Earlier programs nevertheless provide essential
statistical information on UV-bright population.
Figure~\ref{fig:muvcii} (lower panel) illustrates the contribution of new observations to the coverage of [CII]
for UV-bright and UV-faint QSOs.

The CISTERN ([CII]-SelecTed Early universe RefereNce,
  PI: Bouwens) cycle-10 program was designed to obtain [CII]
  observations for essentially all moderately bright
  ($M_{UV,AB}<-21.8$) QSOs and galaxies at $z\geq 5.92$, with a
  published spectroscopic redshift determination, readily observable
  with ALMA (i.e., declination $<$ 30$^\circ$) but lacking
  observations of the [CII] line with ALMA or NOEMA. As targets, 192 sources were
  included in the program, 177 ($>$90\%) of which were
  nominally QSOs.  Of the 192 targets, only 92 were observed with ALMA,  and of these, the data of only 87 could be used.  Multiple targets were
  frequently included in the same scientific goal where possible to
  reduce the cost of the program.

For sources from the CISTERN program, the integration times (per
source) generally ranged from 5 to 20 minutes in duration, with line
and continuum sensitivities of 0.4 mJy/beam and 73$\mu$Jy/beam,
respectively.  The line sensitivity was set to 0.65 mJy/(66 km/s
channel) to allow for the detection (at 5$\sigma$) of [CII] ISM
cooling lines to a limit of $\sim$4-5$\times$10$^{8}$ $L_{\odot}$
(equivalent to an SFR of 40-50 $M_{\odot}$/yr assuming the \citealt{deLooze2014} SFR-L$_{[\text{CII}]}$ relation).  A
synthesized beam of $>$0.9$"$ was requested to avoid over-resolving the
typically extended [CII] emission from bright star-forming galaxies
and QSOs at $z\geq6$.  More details will be provided on the
observations and results for individual sources in the CISTERN survey
paper (R.J. Bouwens et al.\ 2026, in prep).  For sources from other
programs, the integration times were generally longer by about 3 to 4 times, and the sensitivity was twice higher.

Our analysis of the ALMA observations of each target proceeded as
follows.  For each target, measurement sets were created using the
ScriptForPI.py script developed for the QA2 delivery using CASA
\citep{CASA}, and the data were time averaged in 30 s
segments. Finally, the data were reimaged using the task
\textsc{tclean}, adopting natural weighting.  Natural weighting
maximizes the S/N we were able to achieve for line and continuum
detections from CISTERN.  In probing the dust continuum emission from
targets in CISTERN, we excluded the frequency range surrounding
candidate ($>$5$\sigma$) [CII] lines ($\pm$0.4 GHz) to produce
continuum images of the targeted QSOs.

We also used a limited quantity of archival ALMA observations
from the cycle-7 program 2019.1.00074.S (PI: Izumi) targeting the
[CII] line in 12 z$>$5.9 QSOs.  The angular resolution of these
observations was twice higher than that of CISTERN, with line and
continuum sensitivities of $\sim$0.12 mJy/beam (100 km/s channels) and
$\sim$15$\mu$Jy/beam, respectively \citep{Izumi2021,Izumi2021_redQSO},
this is 3.5 times more sensitive than CISTERN. The results for 2 QSOs
from this sample are discussed in detail in \citet{Izumi2021} and
\citet{Izumi2021_redQSO}.

\section{Results}

\subsection{Search for [C II] line emission from z$>$5.9 QSOs}

We performed an analysis of the full set of ALMA data delivered to us
as part of the CISTERN program.  In each data cube, we made use
of our own line-search code to search for $>5\sigma$ line emission from
each of our targets.  Briefly, with this code, we quantified the rms
noise in each spectral channel (which for our delivered data has a
width of $\sim$8 km/s), co-added in quadrature the significance levels
over spectral windows running from 100 km/s to 800 km/s, and then searched
for candidate $>$5$\sigma$ lines in the running significance levels
calculated for each spectral window. Because sources are expected
to be largely unresolved in the observations obtained as part of
CISTERN, with beam sizes $\geq$0.9$''$, it is possible to conduct this
search using peak fluxes (i.e., without any additional spatial
smoothing or tapering of the ALMA observations). Plausible detections
of the [CII] line were made for 62 of the 87 targets from the delivered
CISTERN data for a success rate of 71\%.  Of the 12 targeted QSOs in
cycle-7 program 2019.1.00074.S (PI: Izumi), a [CII] line was identified
in 8 of the QSOs.

\begin{figure*}
\includegraphics[width=18.3cm]{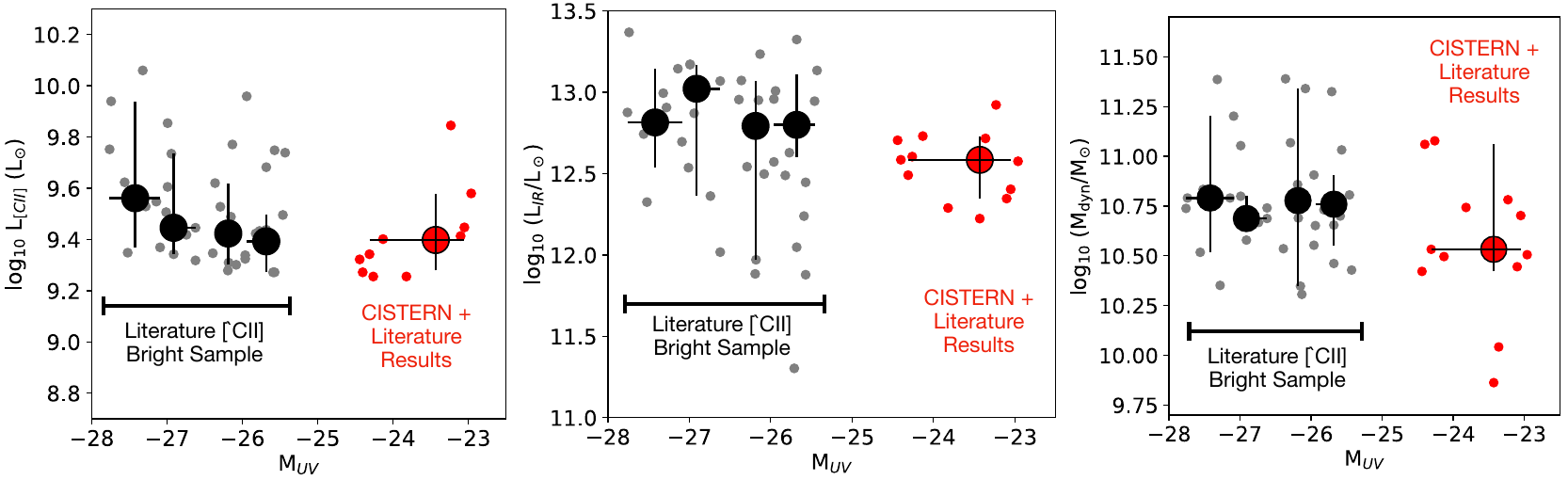}
\vspace{-0.1cm}
\caption{Median [CII] luminosities (\textit{left}), IR luminosities
  ($L_{IR}$; \textit{center}), and [CII] FWHMs (\textit{right}) vs. UV
  luminosity for QSOs in massive host galaxies as inferred from their
  luminous ($L_{\text{[CII]}}$$>$1.8$\times$10$^9$ L$_{\odot}$) [CII]
  emission.  The plotted error bars are 1$\sigma$.  Our $M_{dyn}$
  determinations were derived using the measured FWHMs of the [CII]
  line and a fitting formula from \citet{Neeleman2021}.  Measurements
  for the individual QSOs contributing to the medians for QSOs at
  brighter and fainter UV luminosities are shown with the gray and red
  circles, respectively (from Appendix A and
  Table~\ref{tab:massive_uv_faint}).  The characteristics of UV-bright
  QSOs were derived mostly from the literature, while the
  characteristics of the UV-faint sample were predominantly derived
  from CISTERN.\label{fig:hostgalprop}}
    \end{figure*}

\subsection{Using [C II] to identify QSOs in massive host galaxies over a wide range in
  UV luminosity\label{sec:hostgal}}

We used the [CII] luminosity to segregate our QSOs by host galaxy mass. Because the [CII] luminosity can be used as a measure of the molecular gas mass and SFR, this approach is well motivated.  It is worth noting, however, that in many analyses of QSOs and their host galaxies, the apparent dynamical mass is used as the fiducial mass of the QSO host galaxies \citep[e.g.,][]{Willott2013,Izumi2019,Izumi2021,Neeleman2021}. Because these mass estimates are sensitive to the inclination angle inferred for the host galaxies and the spatial resolution of our ALMA observations is low, we decided to use the [CII] luminosities as a proxy for the host galaxy mass.

We compare the [CII] luminosities we inferred for our QSO sample
(based on the peak fluxes) with their UV luminosities in
Fig.~\ref{fig:muvcii} (top).  $5\sigma$ upper limits were computed for
QSOs without a detected [CII]$_{158\mu\text{m}}$ line assuming a full
width at half maximum (FWHM) of 225 km/s (median FWHM$_{[\text{CII}]}$
of QSOs with $L_{[\text{CII}]}<5\times10^{8}$ $L_{\odot}$ observed by
\citealt{Izumi2018,Izumi2019}) and coverage of
        [CII]$_{158\mu\text{m}}$ with the acquired ALMA band 6 data.
        While there is a weak correlation between the observed [CII]
        luminosity for QSOs and their UV luminosities, as
        \citet{Izumi2019} noted, enormous scatter is present in the
        relation, with a modest fraction of QSOs being both very
        bright in the rest-UV and faint in [CII] or being very bright
        in [CII] and faint in the rest-UV.

The QSOs that are UV-faint 
but are extremely luminous in [CII]are of particular interest.  In a few cases, their [CII] luminosities
rival the [CII] luminosities of UV-bright
($M_{UV,AB}\sim-27$ mag) QSOs.  The the UV luminosity of the most [CII] luminous of these QSOs
(J144045.91+001912.9) is $-23.2$ mag and is brighter than all but three other QSOs shown in our
Table~\ref{tab:datasets} compilation. Because [CII] is a known tracer of the molecular gas mass in host galaxies,
the implication is that these UV-faint QSOs live in similar-mass host
galaxies as UV-bright QSOs.  

To investigate whether the host galaxies of these [CII]-luminous 
UV-faint QSOs really are as massive as QSOs at the bright end of the UV LF, 
we identified a selection of [CII]-luminous QSOs across a range of UV 
luminosity and evaluated whether the physical properties appeared to show any dependence on the UV
luminosities of the QSOs.  For our [CII] selection, we required that sources show $L_{\text{[CII]}}>1.8$$\times$10$^9$ $L_{\odot}$.  As 1.8$\times$10$^9$ $L_{\odot}$ is the median luminosity of QSOs in our UV-bright ($M_{UV,AB}<-26.0$) subsample, the purpose of this selection was to identify QSOs whose host galaxies were at the high-mass end of the QSO population at high UV luminosities. Of the 190 QSOs considered as part of this analysis, 63 qualified as having high-mass host galaxies based on this definition.  This [CII] luminosity (1.8$\times$10$^9$ $L_{\odot}$) also corresponds to host galaxy gas masses of $\sim$10$^{10.5}$ $M_{\odot}$.  

These [CII] luminous QSOs were then segregated into four brighter UV luminosity bins
and one faint bin, that is, (1) $-28.0<M_{UV,AB}<-27.0$, (2)
$-27.0<M_{UV,AB}<-26.5$, (3) $-26.5<M_{UV,AB}<-26.0$, (4)
$-26.0<M_{UV,AB}<-24.5$, and (5) $-24.5<M_{UV,AB}<-21.5$.  The number of [CII]-bright QSOs in each UV luminosity bin is 9, 14, 8, 19, and 13 out of 11, 29, 21, 67, and 61 QSOs, respectively, resulting in a [CII]-luminous fraction of 0.82, 0.48, 0.38, 0.27, and 0.23, respectively, for these different UV-luminosity bins. The monotonic decrease from brighter to fainter UV luminosities is clear.  The
13 UV-faint QSOs that we discovered to be [CII] luminous by leveraging the new CISTERN data and from archival observations or the literature are presented in Table~\ref{tab:massive_uv_faint}.

For the QSOs in each UV luminosity bin, we quantified three different
characteristics of the QSOs that were to provide an approximate probe
of the overall mass of their host galaxies: (1) their [CII] luminosity
$L_{\text{[CII]}}$, which is a known tracer of the host galaxy gas
mass,\footnote{Even though the [CII] luminosity was used to construct
the QSO subsamples for this exercise, we decided to include the [CII]
luminosity here along with the other two measures of mass to allow for
side-by-side comparisons.} (2) the IR luminosity $L_{\text{IR}}$ of
these sources, and (3) the FWHM of the [CII] line, which correlates
with the dynamical mass of the host galaxies.  We computed the IR
luminosity (8 - 1000$\mu$m) of QSOs using the same method and
assumptions as \citet{Venemans2020}, following the \citet{daCunha2013}
treatment to account for the effect of the cosmic microwave background
on the measured fluxes. The inferred IR luminosities clearly depend on
the adopted far-IR spectral energy distribution, and individual QSOs
exhibit some diversity in their far-IR spectral shapes
\citep[e.g.,][]{Lyu2017}.  If the far-IR spectral energy distribution
of QSOs showed a higher or lower dust temperature than the fiducial
$T_d = 47$ K adopted by \citet{Venemans2020}, we might expect
2.3$\times$ higher IR luminosities, for example, if the dust
temperatures were $T_d \sim 60$ K (upper end range seen by
\citealt{Leipski2013}) or 30\% lower if the dust temperatures were
closer to $T_d \sim 41$ K \citep{Priddey2001}.

We present the median $L_{\text{[CII]}}$, $L_{\text{IR}}$, and
FWHM$_{[\text{CII}]}$ derived for sources in each UV luminosity bin
in the left, center, and right panels, respectively, of
Fig.~\ref{fig:hostgalprop}.  Our $M_{dyn}$ determinations were
derived using the measured FWHMs of the [CII] line and a fitting
formula from \citet{Neeleman2021},
\begin{equation}
  M_{dyn} = (4.0_{-2.0}^{+4.0} (_{-2.8}^{+2.4})) \left (\frac{\text{FWHM}}{\text{km}\,\text{s}^{-1}}\right )^2 \times 10^5 M_{\odot} .
\end{equation}
Because of the differential comparison we made, we expected no effect
from the overall scale factor on $M_{dyn}$.  While this relation was
derived from UV-bright QSOs, we expect its accuracy to depend on the
characteristics of the host galaxy itself and not on the rest-UV
luminosity of the QSO. Even though this assumption is not fully accurate,
the similar [CII] FWHM distributions of [CII]-bright QSOs spanning a
wide range in UV luminosity suggest broadly similar dynamical
properties.

The determinations for the brighter sample and faint sample are shown with
solid black circles and solid red circle, respectively.  For each of
the three characteristics, the median derived for the faint sample is
consistent with that found for the brighter samples, with the possible
exception of the IR luminosities computed for the QSOs where a contribution from active galactic nucleus heating for the brighter QSOs is expected
\citep{Schneider2015,Kirkpatrick2015}.  The broad consistency between
their [CII] luminosities (by construction through the selection
criteria) and $M_{dyn}$ suggests a broad similarity in the host galaxy
masses for each of these samples.

    \begin{figure}
\includegraphics[width=9.1cm]{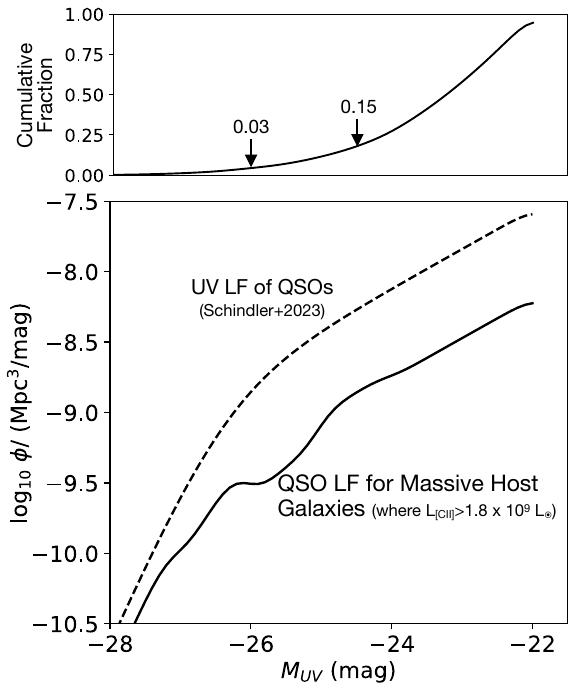}
\caption{\textit{Top}: Fraction (vertical axis) of massive host galaxies (with $L_{\text{[CII]}}>1.8\times10^{9}$ $L_{\odot}$: solid line) brighter than some UV luminosity (horizontal axis).  Our analysis indicates that the UV luminosities of only 15\% and 3\% of the QSOs in massive host galaxies are brighter than $-$24.5 mag and $-$26.0 mag, respectively (see Sect. \ref{sec:muvdist}).  The vast majority of the QSOs are UV-faint, i.e., $>$$-$24.5 mag.  The cumulative fraction presented here is based on a consideration of QSOs brightward of $-$22 mag where broad [CII] coverage of QSO samples exists.  \textit{Bottom}: Volume density of QSOs with massive host galaxies vs. UV luminosity (solid line).  The dashed line represents the \citet{Schindler2023} LF and is shown for comparison. While the percentage of UV-bright QSOs with massive host galaxies (48-82\%) is much $\sim$3-4$\times$ higher than for the UV-faint ($M_{UV,AB}>-24.5$) QSOs (21-28\%), $>$85\% of the QSOs with massive host galaxies are still UV-faint.\label{fig:muvdist}}
\end{figure}

   \begin{figure}
   \includegraphics[width=9.1cm]{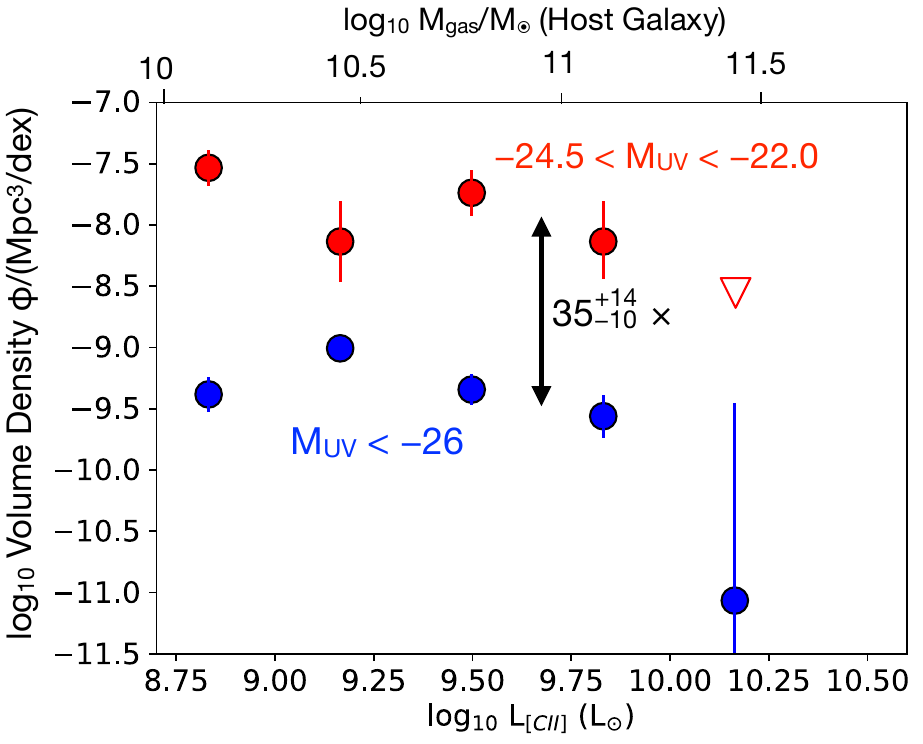}
   \caption{[CII] LFs inferred for UV-faint ($-24.5<M_{UV,AB}<-22.0$:
     red circles) and UV-bright QSOs ($M_{UV,AB}<-26.0$: solid blue
     circles) at z$\sim$6 based on the distribution of [CII]
     luminosities seen for QSOs at a given UV luminosity and the
     \citet{Schindler2023} $z\sim6$ UV LF.  The uncertainties were
     computed by adding in quadrature the contribution from every QSO
     whose measured [CII] luminosity corresponded to a given bin.  The
     downward-pointing triangle is a 1$\sigma$ upper limit.  For
     $z\sim7$ QSOs \citep{Matsuoka2023}, both LFs would be
     $\sim$5$\times$ lower (see Sect. \ref{sec:muvdist} for details).
     The upper axis indicates the approximate host galaxy gas mass
     adopting the [CII] conversion factor from \citet{Zanella2018}.
     The volume density of [CII] luminous galaxies that are UV-faint
     is $\sim$35$_{-10}^{+14}$$\times$ higher than [CII] luminous
     galaxies that are UV-bright.  Significant uncertainties remain as
     to what (dust, low $M_{BH}$, or extreme sub-Eddington accretion)
     drives the differences between UV-bright and UV-faint QSOs in
     massive host galaxies (Sect. 4).  \label{fig:ciilf}}
    \end{figure}

\subsection{UV luminosity distribution of QSOs in massive host galaxies\label{sec:muvdist}}

The substantial [CII] coverage we have of UV-bright QSOs
and now UV-faint QSOs means that we can provide a first quantification of the
UV luminosity distribution of massive host galaxies.  As in the
previous section, we assumed that QSOs live in a massive host galaxy when their
[CII] luminosity is greater than 1.8$\times$10$^{9}$ $L_{\odot}$.

To perform this exercise, we started with the QSO UV LF results of
\citet{Schindler2023} at $z\sim6$.  We then broke the UV LF
into five different bins with an UV luminosity $-28<M_{UV,AB}<-27$,
$-27<M_{UV,AB}<-26$, $-26<M_{UV,AB}<-25$, $-25<M_{UV,AB}<-24$, and
$-24<M_{UV,AB}<-22$ and then multiplied the LF at that UV luminosity
by the fraction of sources with [CII]$_{158\mu\text{m}}$ luminosities in excess of
1.8$\times$10$^{9}$ $L_{\odot}$.  We present the result in the lower panel
of Fig.~\ref{fig:muvdist} to $M_{UV,AB}\sim-22$ , where CISTERN
provides good coverage of [CII]$_{158\mu\text{m}}$.  The step-like structure in the QSO LF we inferred for massive host galaxies derives from the [CII] distribution being split across distinct UV-luminosity bins. The figure clearly shows that the vast majority of QSOs in massive host galaxies appear to
be UV-faint.

To better quantify the fraction of QSOs in massive host galaxies that are UV-bright or UV-faint, we plot the cumulative fraction
above a given UV luminosity in the upper panel of
Fig.~\ref{fig:muvdist}.  Only 15\% of the QSOs lie brightward of 
$-$24.5 mag, and just 3\% lie brightward of $-$26 mag, meaning that 
the vast majority ($>$85\%) of QSOs in massive host galaxies are UV-faint, that is, $M_{UV,AB}>-24.5$ mag. We only considered the fraction to a UV luminosity of $-22$ mag because our sampling of the [CII] distribution for fainter QSOs only extends to that UV luminosity.

In addition to our quantifying the UV luminosity distribution of QSOs
in massive host galaxies, it is interesting to derive the volume
density of UV-bright and UV-faint QSOs as a function of their host
galaxy gas masses. When we continue to treat $L_{[\text{CII}]}$ as a proxy for the host galaxy mass, this can be done by deriving the [CII] LF for UV-bright and UV-faint QSOs.  
    
As in our earlier determination of the UV luminosity distribution for
QSOs with massive host galaxies, we again broke the UV LF into five
different bins in UV luminosity.  We then reassigned galaxies in each
UV LF bin to QSOs with a given [CII] luminosity in accordance with the
observed $L_{\text{[CII]}}$ distribution in that UV luminosity bin. We
effectively made the assumption that specific z$>$6 QSOs were only
targeted on the basis of their UV luminosity and were not affected by
their characteristics in the far-IR (i.e., $L_{[CII]}$ or $L_{IR}$).
In computing the uncertainties on the [CII] LF, the contribution from
every QSO with a [CII] luminosity measurement is added in quadrature.
LF bins with larger uncertainties on the volume density are based on a
smaller number of QSOs with that luminosity measurement.

We present the results for a UV-bright ($M_{UV,AB}<-26.0$) and
UV-faint ($-24.5<M_{UV,AB}<-22.0$) QSO selection in
Fig.~\ref{fig:ciilf} and Table~\ref{tab:binnedlfs}.
Over a broad range of [CII]$_{158\mu\text{m}}$
  luminosities (or equivalently, host galaxy masses), the volume
  density of UV-faint QSOs exceeds that in UV-bright QSOs by a
  factor of $\sim$20-40.  For [CII] luminosities in excess of
1.8$\times$10$^{9}$ $L_{\odot}$, the excess is 35$_{-10}^{+14}$.
Similar excesses would be obtained if the comparison were made using
other thresholds in [CII]$_{158\mu\text{m}}$ luminosity.  The
sensitivity of the ALMA data available to CISTERN limits our ability
to probe this relation below $L_{\text{[CII]}} \sim 10^{8.5}$
$L_{\odot}$.

We found a similar excess of [CII]-bright UV-faint QSOs relative to
UV-bright systems using other $z\sim6$ QSO LF determinations.  For
the \citet{Matsuoka2018_QSOLF} and \citet{Willott2010_QSOLF} LFs, the
excess of UV-faint [CII]-bright systems over bright systems is
26$_{-8}^{+11}$ and 38$_{-11}^{+16}$, respectively, which is consistent
($\pm$0.1 dex) of that found using the \citet{Schindler2023} LF.  When
we instead relied on the UV LF of QSOs at z$\sim$7
\citep{Matsuoka2023}, the excess would be 20$_{-6}^{+8}$, with an about five times lower normalization.

Results from this section strongly suggest that the vast majority
($>$85\%) of QSOs with massive host galaxies are UV-faint, that is,
with $M_{UV,AB}>-24.5$ mag.  By contrast, only a small fraction of
massive host galaxies ($<$3\%) contain QSOs that are as bright as
those discovered in early QSO work using the Sloan Digital Sky Survey
\citep[e.g.,][]{Fan2001}.

\begin{table} 
\centering
\caption{Binned [CII] LF results for QSOs at $z\sim6$$^a$.}
\label{tab:binnedlfs}
\begin{tabular}{c c}
\hline
log$_{10}$ (L$_{[\text{CII}]}$$/$L$_{\odot}$) & $\log_{10} (\phi^* \, \text{dex}^{-1} \text{Mpc}^{-3})$\\
\hline\hline
\multicolumn{2}{c}{Bright ($M_{UV,AB}<-26$) QSOs}\\
\hline
8.83 & $-$9.38$\pm$0.14 \\
9.17 & $-$9.01$\pm$0.09 \\
9.50 & $-$9.34$\pm$0.13 \\
9.83 & $-$9.56$\pm$0.18 \\
10.16 & $-$11.07$\pm$1.62 \\\\
\multicolumn{2}{c}{Faint ($-24.5<M_{UV,AB}<-22$) QSOs}\\
\hline
8.83 & $-$7.53$\pm$0.15 \\
9.17 & $-$8.13$\pm$0.33 \\
9.50 & $-$7.74$\pm$0.19 \\
9.83 & $-$8.13$\pm$0.33 \\
10.16 & $<$$-$8.38$^b$\\
\hline\hline
\end{tabular}\\
\tablefoottext{a}{[CII] LF results shown in Fig.~\ref{fig:ciilf}.  The results were inferred using the $z\sim6$ QSO LF from \citet{Schindler2023} and the observed $L_{[\text{CII}]}$ vs. $M_{UV}$ distribution shown in Fig.~\ref{fig:muvcii}. The results here are the same as in Fig.~\ref{fig:ciilf}.}

\tablefoottext{b}{$1\sigma$ Upper Limit}
\end{table}

\section{Discussion}

\subsection{UV-faint QSOs in massive host galaxies at $z\sim6$: Difference from UV-bright QSOs\label{sec:gen_discussion}}

In the previous section, we demonstrated that the vast majority
($>$85\%) of QSOs with massive host galaxies are UV-faint, that is, $M_{UV,AB}>-24.5$ mag, which is much fainter than the initial QSOs discovered at z$\sim$6 \citep{Fan2006}.  As UV-faint QSOs represent the
dominant form of QSO in massive host galaxies, it is essential that we
ascertain their nature and how they differ from UV-bright
QSOs at z$\sim$6-7 that have been the focus of study for the past two decades.

   \begin{figure*}
\includegraphics[width=17.5cm]{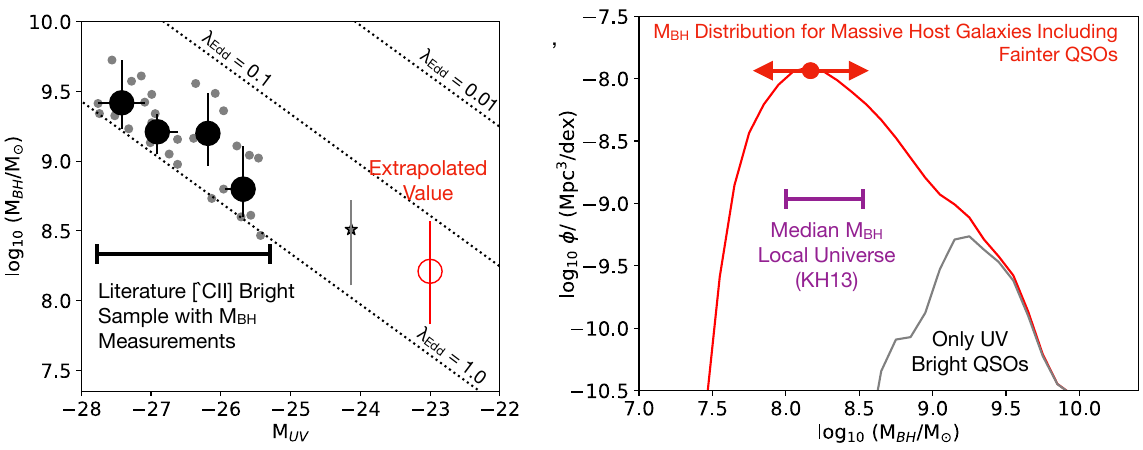}
\caption{\textit{Left}: Median measured $M_{BH}$ (solid black
    circles) from the literature for QSOs in massive host galaxies
  with $L_{\text{[CII]}}$$>$1.8$\times$10$^9$ L$_{\odot}$ vs. UV
  luminosity.  The smaller solid gray points show the measured $M_{BH}$
  vs. $M_{UV}$ for the massive host galaxy sample.  The $M_{BH}$
  measurements for an $M_{UV,AB}\sim-24.1$ QSO at $z=7.07$
  \citep{Izumi2021} is shown with the gray star.  The dotted diagonal lines indicate $M_{BH}$s corresponding to Eddington ratios $\lambda_{Edd}$ of 1, 0.1, and 0.01.  There is a strongly
  suggestive correlation between the BH masses of QSOs in
  massive host galaxies and their UV luminosities, suggesting that
  UV-brightness of QSOs in massive host galaxies is largely driven by
  the BH mass.  The open red circle shows the extrapolated
  $M_{BH}$ at $M_{UV,AB}\sim-23.0$ mag, together with the $1\sigma$
  uncertainty based on the extrapolation.  \textit{Right}:
  Distribution of $M_{BH}$ in massive host galaxies at $z>6$ in
  UV-bright ($M_{UV,AB}<-26$) QSOs (black line) and including
  QSOs down to a UV luminosity of $-$22.0 mag (red line).
  The precise mass where the $M_{BH}$ distribution peaks has a 1$\sigma$
  uncertainty of $\pm$0.4 dex (red arrows) and is sensitive to the $M_{UV,AB}=-22$ cutoff the QSO LF we assumed and also to the best-fit slope of the $M_{BH}$ vs. $M_{UV}$ relation.  The purple bracketed region indicates the approximate
  range of $M_{BH}$ for QSOs in similar-mass host galaxies in the
  local Universe (e.g., \citealt{Kormendy2013}: KH13).  The median $M_{BH}$ for
  QSOs in massive host galaxies at $z\sim6$ appears to be consistent
  with the local value (see Sect. \ref{sec:mbhdist}), but the uncertainties 
  are too large to be sure.  Further measurements
  of $M_{BH}$ for the UV-faint QSOs in massive host galaxies at
  $z\sim6$ are required to resolve this open
  question.\label{fig:mbhdist}}
    \end{figure*}

   There are three different physical explanations for why the
   majority of QSOs in massive host galaxies are UV-faint rather
   than UV-bright: (1) dust obscuration, (2) lower BH masses, and
   (3) extreme sub-Eddington accretion.  For the general QSO
   population, the
   latter two possibilities have been discussed at length, with analyses suggesting a shift to lower
   BH masses for UV-faint QSOs \citep{Willott2010,Silverman2025} and
   other analyses suggesting a shift to a lower Eddington ratio
   \citep{Wu2022}, but essentially, all analyses indicated a large
   spread in the BH and host galaxy masses and Eddington ratios
   \citep{Izumi2018,Kim2018,Izumi2019,Onoue2019,Ding2025}.

   We will discuss the possibility of dust obscuration as part of a
   separate analysis that discusses rest-UV colors
   (power-law slopes $\alpha_{\lambda}$) of the fainter QSOs in detail using
   various near-IR ground-based imaging and WISE observations, but
   this only appears to be the case for $\sim$25\% of the QSOs in our
   [CII]-bright but UV-faint selection (R.J. Bouwens et al.\ 2026,
   in prep).\footnote{This is only a preliminary estimate
   based on the derived slopes $\alpha_{\lambda}$ from the available
   near-IR and WISE imaging data.  For a more definitive estimate, it
   might be valuable to obtain more sensitive observations to redder
   wavelengths (e.g., JWST/MIRI imaging observations), which
   might reveal a more substantial obscured component
   \citep[e.g.,][]{Lyu2024}.}  Two of the clearest cases of dust
   obscuration in our CISTERN + literature selection, that is,
   HSCJ114632$-$015438 and J1205$-$0000, were discussed
   by \citet{Kato2020} and \citet{Izumi2021_redQSO}, respectively.
   A third QSO, J144045.91+001912.9, also shows clear
   evidence of dust reddening based on its near-IR colors and,
   even more convincingly, a recent rest-UV spectrum obtained with
   JWST/NIRSpec in Cycle 4 (D. Yang et al.\ 2026, in prep).

To evaluate whether one of the two other potential explanations of
why most QSOs in massive host galaxies are UV-faint, measurements of
the BH masses for the UV-faint QSOs in our CISTERN+archival sample
are needed.  Unfortunately, only two QSOs satisfying our constraints
of living in a massive host galaxy (i.e.,
$L_{\text{[CII]}}>1.8\times10^{9} L_{\odot}$) have such measurements,
that is, J1205$-$0000 and J1243+0100, and J1205$-$0000 seems to be most
consistent with the dust-obscuration hypothesis.\footnote{Using photometry
in WISE, \citet{Kato2020} suggested that J1205$-$0000 might be a partially
reddened QSO.  Applying an extinction correction to the results of
\citet{Onoue2019}, \citet{Kato2020} reported $M_{BH} = 2.9_{-0.8}^{+0.3}
\times10^{9} M_{\odot}$ and $\lambda_{Edd} = L_{bol}/L_{Edd} =
0.22_{-0.03}^{+0.04}$.  These $M_{BH}$ results are consistent with
what has been found for UV-bright QSOs at $z\sim6$
\citep{Kurk2007,Jiang2007,Willott2010,DeRosa2011,DeRosa2014,Venemans2015}
although the accretion efficiency $\lambda_{Edd}$ is twice lower,
which agrees with lower-redshift QSO results as reported by
\citet{Shen2011}.} For J1243+0100, where $M_{UV,AB}=-24.13$ mag and
$L_{\text{[CII]}}=2.5\times10^9 L_{\odot}$,
\citet{Matsuoka2019_faintz7qso} derived $M_{BH}=(3.3\pm2.0)\times10^{8}
M_{\odot}$ from spectroscopy of Mg$\,\textsc{ii}\,2800\,\AA$ while
inferring $\lambda_{Edd}$ to be equal to $0.34\pm0.20$.  The mass of
the BH in J1243+0100 \citep{Matsuoka2019_faintz7qso} is
lower by about  ten times than for UV-bright QSOs at $z\sim6$. The accretion
efficiency is also twice lower.  This suggests that the about ten times lower BH mass of
J1243+0100 appears to be the most
important factor driving its faintness in the rest-UV relative to
UV-bright QSOs at z$\sim$6.

If the primary driver of the variations in the UV luminosity of QSOs is
instead extreme sub-Eddington accretion or dust obscuration (as is
plausibly the case for HSCJ114632-015438, J1205-0000, J0958+0139, and
J144045.91+001912.9:
\citealt{Kato2020,Izumi2021_redQSO,Endsley2023_CII}; D. Yang et
al. 2026, in prep) and not $M_{BH}$, this would mean that
UV-faint QSOs might then just correspond to different phases in the
lifetime of UV-bright QSOs, as implied by the relatively short QSO
lifetimes measured for most z$>$6 QSOs
\citep{Davies2019,Eilers2020,Eilers2021} and halo masses inferred for
QSOs \citep{Shen2007,Eilers2024,Pizzati2024}.

Interestingly enough, in the scenario in which dust obscuration or
sub-Eddington accretion are the primary drivers of variations in the UV
luminosity, the black mass $M_{BH}$ in UV-faint [CII]-bright QSOs
would be similar to that of the UV-bright QSO population, that is,
$M_{BH}\gtrsim 10^9$ $M_{\odot}$. Based on the $\sim$30 times higher
prevalence of UV-faint [CII]-bright QSOs relative to the
UV-bright QSOs, this would imply a correspondingly larger population
of $M_{BH}\sim10^9$ $M_{\odot}$ BHs at $z\sim6$-7 than exists
in the UV-bright QSO population (by $\sim$30 times), making the
assembly of the $M_{BH}\sim10^9$ $M_{\odot}$ SMBHs population by
$z\sim6$-7 even more challenging
\citep[e.g.,][]{Volonteri2012,Inayoshi2020,Fan2023}.

\subsection{QSOs in massive host galaxies: Dependence of $M_{BH}$ and $\lambda_{Edd}$ on $M_{UV}$\label{sec:muvdep}}

For the remaining sources, we do not yet have spectroscopy probing
their BH masses to ascertain whether their faintness in the UV is the
result of (1) their BH masses being lower than the UV-bright QSO
population or (2) accretion efficiencies $\lambda_{Edd}$ being
lower. To understand the dominant UV-faint QSO population in massive
host galaxies, we focused on QSOs in massive host galaxies (which we
take to mean $L_{\text{[CII]}}>1.8\times10^9$ $L_{\odot}$) with
measured $M_{BH}$ and $\lambda_{Edd}$.  34 [CII]-luminous QSOs appear
to have these measurements available and extend from $-29.09$ mag to
$-24.13$ mag, with the majority in the range $-28<M_{UV,AB}<-25$. The
results for these QSOs are included in
Table~\ref{tab:mbhsample}. These are approximately the same sets of
QSOs that we considered in Sect. \ref{sec:hostgal} to search for
consistency in the apparent host galaxy masses versus a function of UV
luminosity.

\begin{figure*}
\includegraphics[width=17.5cm]{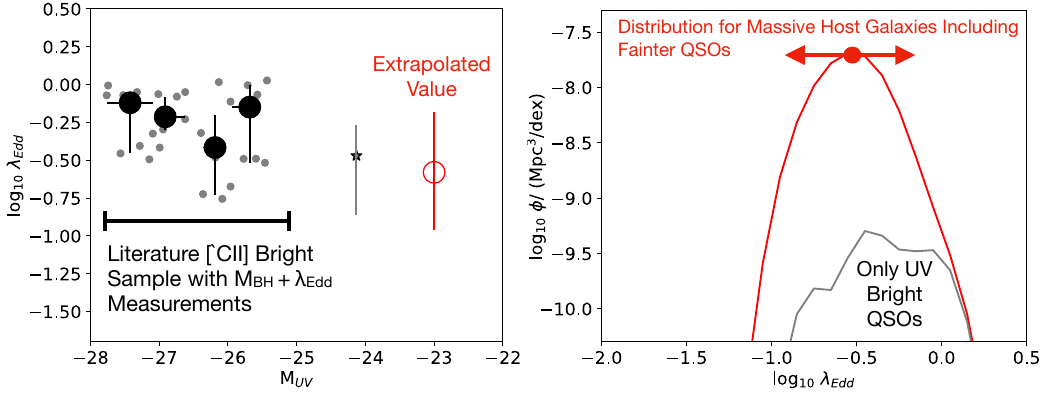}
\caption{\textit{Left}: Median inferred $\lambda_{Edd}$ (solid black circles)
  from the literature for QSOs in massive host galaxies with
  $L_{\text{[CII]}}$$>$1.8$\times$10$^9$ L$_{\odot}$ vs. UV
  luminosity.  The gray points and gray star are the same as in
  Fig.~\ref{fig:mbhdist}.  No strong correlation is seen between the
  UV luminosity of QSOs in massive host galaxies and their accretion
  efficiency $\lambda_{Edd}$.  As such, it would appear that $M_{BH}$
  is the dominant factor impacting the UV-brightness (or $L_{bol}$)
  of $z\sim6$-7 QSOs and the duty cycle (i.e., variations in
  $\lambda_{Edd}$) have a more limited impact.  The open red circle
  shows the extrapolated $M_{BH}$ at $-$23.0 mag, together with the
  $1\sigma$ uncertainty based on the extrapolation.  \textit{Right}:
  Distribution of $\lambda_{Edd}$ in massive host galaxies at $z>6$ in
  UV-bright ($M_{UV,AB}<-26$) QSOs (black line) and including QSOs
  down to a UV luminosity of $-$22.0 mag (red line).  Given the
  large uncertainties in the precise dependence $\lambda_{Edd}$ shows
  on UV luminosity, the peak of the distribution is very uncertain
  ($\pm$0.4 dex: red arrows indicate the 1$\sigma$
  uncertainties).\label{fig:erdist}}
    \end{figure*}

For each subsample, we computed the median $M_{BH}$, $\lambda_{Edd}$,
and $M_{UV}$ for sources in the selection.  For our estimates of
$\lambda_{Edd}$ for individual QSOs, we used the observed UV
luminosities (1450\AA) to estimate $L_{bol}$ using the equations
presented by \citet{Runnoe2012} and then took $\lambda_{Edd} =
L_{bol}/L_{Edd}$.  The results are shown in the left panels of
Figs.~\ref{fig:mbhdist} and \ref{fig:erdist}.  Interestingly enough,
no sources are found with $\lambda_{Edd}$ substantially in excess of
1, similar to results from other analyses
\citep[e.g.,][]{Farina2022,Kelly2013}.

In the case of $M_{BH}$, we found a clear trend
from higher $M_{BH}$ to lower $M_{BH}$ masses faintward
in UV luminosity. When we adopted a simple power-law relation in
modeling in the $M_{BH}$ versus $M_{UV}$ results (also including the \citealt{Izumi2021} z=7.075 source), we recovered the following best-fit relation:
\begin{equation}
\log_{10} (M_{BH} / M_{\odot}) = (9.27_{-0.10}^{+0.11}) + (-0.26_{-0.11}^{+0.10})(M_{UV}+27).
\end{equation}Based on this fitting formula, the expected median $M_{BH}$ for our sample of UV-faint QSOs in massive host galaxies is shown in Fig.~\ref{fig:mbhdist} with the open red circle (left panel).

We note that evidence for such a correlation between the UV
luminosity and $M_{BH}$ was the subject of extensive
discussion in a number of earlier
\citep[e.g.,][]{Willott2010,Izumi2018,Onoue2019,Li2022} and
contemporary \citep{Silverman2025} studies. Our analysis is different in the consistent application of similar host galaxy
selection criteria to quantify trends in UV luminosity.

No clear trend is evident in the accretion efficiency $\log_{10}
\lambda_{Edd}$ versus UV luminosity (Fig.~\ref{fig:erdist}).  Again assuming a simple power-law
relation in modeling the $\lambda_{Edd}$ versus $M_{UV}$ results, we
found the following best-fit relation:
\begin{equation}
\log_{10} \lambda_{Edd} = (-0.21_{-0.09}^{+0.09}) + (-0.09_{-0.10}^{+0.11})(M_{UV}+27).
\end{equation}
Based on this fitting formula, the expected median $\lambda_{Edd}$ for
our sample of UV-faint QSOs in massive host galaxies is shown in
Fig.~\ref{fig:erdist} with the open red circle (left panel).

Similar to our discovery of a correlation between $M_{BH}$ and UV
luminosity in massive host galaxies, earlier studies have also
reported a limited correlation between the accretion
efficiency $\lambda_{Edd}$ of QSOs and their $M_{UV}$
\citep[e.g.,][]{Willott2010_BHMF,Farina2022} for QSOs at UV
luminosities of $-$23 mag and brightward.  This is consistent with the
idea that SMBH growth occurs through accretion at near the Eddington
rates or at extreme sub-Eddington rates
\citep{Narayan1998,Kondapally2025} (and thus cannot be identified as a
z$>$5.9 QSO).\footnote{For some z$\sim$6 QSOs, quenching
might already be occurring \citep{Onoue2024}.} Based on
  QSO lifetime measurements, SMBHs likely spend part of the build-up
  phase in a more obscured state
  \citep[e.g.,][]{Eilers2017,Davies2019}, and there are also likely
  significant periods of time when accretion onto the SMBH
  slows dramatically.

\subsection{Distribution of $M_{BH}$ and $\lambda_{Edd}$ for QSOs in massive host galaxies at z$\sim$6-7\label{sec:mbhdist}}

It is interesting to use the results derived in the previous section
to try to estimate the approximate distribution in $M_{BH}$ and
$\lambda_{Edd}$ for QSOs in massive host galaxies and then to compare
the derived distribution with that found in the local Universe.  Given
the apparent dependence of the measured $M_{BH}$ on the UV luminosity,
it is possible that the inclusion of fainter QSOs might have a
sizeable impact on the median $M_{BH}$ inferred for QSOs in massive
host galaxies.

For this exercise, we considered the UV luminosity distribution of QSOs
in massive host galaxies that we present in Fig.~\ref{fig:muvdist}.
For each 0.1 mag bin in UV luminosity, we distributed QSOs into
different bins of $M_{BH}$ and $\lambda_{Edd}$ according to the
empirical relation presented in the left panels of
Figs.~\ref{fig:mbhdist} and \ref{fig:erdist}. To construct the
$M_{BH}$ and $\lambda_{Edd}$ for QSOs brighter than $-25$ mag, we used the empirical distribution seen in the observations (Appendix
A), but for QSOs with UV luminosities fainter than $-25$ mag, we
shifted the observed distribution of $M_{BH}$ and $\lambda_{Edd}$ to
lower luminosities adopting the best-fit trends derived in
Sect. \ref{sec:muvdep}.

We present the results in the right panels of Figs.~\ref{fig:mbhdist}
and \ref{fig:erdist}.  The black curves only include those UV-bright QSOs in massive host galaxies, while the red curves
include QSOs above a UV luminosity of $-22$ mag (where there is a
reasonable sampling of $L_{\text{[CII]}}$ distribution).  For context, we show in
Fig.~\ref{fig:mbhdist} the approximate range of $M_{BH}$
(purple bracketed region) we would expect for QSOs in similar
mass host galaxies in the local Universe. To derive this range
(10$^{8.1}$ to 10$^{8.5}$ $M_{\odot}$), we used the
\citet{Kormendy2013} $M_{BH}$-$M_{dyn}$ relation and the median
dynamical mass reported by \citet{Neeleman2021} for [CII]-luminous
($L_{\text{[CII]}}>1.8\times10^9$ $L_{\odot}$) QSOs, that is,
$\sim$4$\times$10$^{10}$ $M_{\odot}$.

The distribution in $M_{BH}$ we derived for QSOs in massive
host galaxies at z$\sim$6-7 peaks at a mass of $\sim$10$^{8.1}$
$M_{\odot}$.  The distribution in $M_{BH}$ we estimated for
QSOs in massive galaxies is consistent with the local relation, but lower by a
factor of 15$_{-9}^{+25}$  in mass than found for UV-bright
QSOs \citep{Shen2019,Yang2021,Farina2022}, suggesting that the first
UV-bright QSOs discoveries at z$\sim$6
(e.g., \citealt{Fan2001}) were simply extreme high-mass outliers in
$M_{BH}$ versus $M_{host,gal}$ space.

Significant discussion
\citep[e.g.,][]{Willott2005,Fine2006,Lauer2007,Li2022} has suggested that
UV-bright QSOs might simply be high-mass outliers on the $M_{BH}$
versus $M_{host,gal}$ relation and that the evolution
in this relation can be more reliably assessed by focusing on host galaxies with lower-mass BHs.  In particular, \citet{Willott2013,Willott2015,Willott2017_faintQSO} used near-IR spectroscopy and dynamical information from ALMA
observations for a small sample of UV-faint QSOs and reported suggestive
evidence that UV-faint QSOs showed a $M_{BH}$ versus $M_{dyn}$ relation
that is completely consistent with that in the local Universe
\citep{Kormendy2013}.  Further support for this idea came from
observations of a larger sample of UV-faint QSOs at
$z\sim6$-7 by \citet{Izumi2018}, \citet{Onoue2019}, and
\citet{Izumi2021} and now also leveraging observations from a JWST/NIRSpec program \citep{Onoue2021,Ding2023,Ding2025,Silverman2025}.

The distribution in the accretion efficiency $\lambda_{Edd}$ 
peaks at 10$^{-0.6}$, which is slightly lower than the value for UV-luminous galaxies, but is still broadly consistent, suggesting that the duty cycle only has a limited effect on the composition of QSO samples over a range of UV luminosities (see, however, \citealt{Wu2022}). Based on our results, it would instead appear that $M_{BH}$ is the dominant factor affecting the UV-brightness (or $L_{bol}$) of $z\sim6$-7 QSOs, as was found previously \citep[e.g.,][]{Willott2010}.

We emphasize that there are large uncertainties in the $M_{BH}$ and
$\lambda_{Edd}$ distributions we derived. For the dependence of
$M_{BH}$ on $M_{UV}$ higher or lower by  $1\sigma$ than the best-fit
trend, the peak in $M_{BH}$ for the distribution shifts higher or
lower by 0.4 dex, and similarly for the peak in $\lambda_{Edd}$ when we
propagate the uncertainties.

Based on the current uncertainties, it is essential that
spectroscopic campaigns on UV-faint QSOs at z$\geq$6 continue, so that
$M_{BH}$ and $\lambda_{Edd}$ can be derived for a much larger number of QSOs.  While progress on this front is already being made
\citep[e.g.,][]{Ding2023,Onoue2024} with JWST and in the near future with the substantial Aether dataset \citep{Farina-aether}, it is especially valuable to obtain these measurements for UV-faint QSOs in massive host galaxies.  

These same host galaxies might feature UV-faint and UV-bright QSOs at different points in their life cycles, and it is therefore clear that it should be a priority to focus more on the UV-faint phase with JWST to match the resources committed to the study of UV-bright QSOs better.  While some UV-faint QSOs will be so due to their lower mass BHs, others will be so due to variations in their accretion efficiency or obscuration, as implied by recent results on QSO lifetime and clustering \citep[e.g.,][]{Eilers2020,Eilers2021,Eilers2024,Pizzati2024}.

\section{Summary}

We used observations of the [CII]$_{158\mu\text{m}}$ ISM cooling line to segregate QSOs as a function of mass of the host galaxy, with the aim of identifying particularly massive host galaxies.  The [CII] luminosities of the QSOs, a known tracer of the molecular gas mass, was treated as a proxy for the host galaxy mass and was used to sort 190 QSOs at z$>$5.9, spanning a range of 6 mag in UV luminosity ($-28<M_{UV,AB}<-22$).  The [CII] coverage from a cycle-10 program CISTERN of 46 UV-faint ($M_{UV,AB}>-24.5$) QSOs and 25 especially UV-faint ($M_{UV,AB}>-23.5$) QSOs was particularly valuable for this enterprise. It improved the statistics by five and six times, respectively.

Taking QSOs with [CII] luminosities in excess of 1.8$\times$10$^{9}$ $L_{\odot}$ as having a massive host galaxy, we found 63 such QSOs, including 13 and 7 that are UV-faint ($M_{UV,AB}>-24.5$) and especially UV-faint ($M_{UV,AB}>-23.5$).  The bulk (9/13) of the UV-faint QSOs discovered to be bright in [CII] came from the new CISTERN program.  Remarkably, the dynamical masses $M_{dyn}$ of these UV-faint QSOs were also similar to those of UV-bright QSOs, and their estimated IR luminosities are only slightly fainter than those of UV-bright [CII]-bright QSOs, strongly suggesting that their host galaxy masses are similar to those of UV-bright QSOs.

Using these selections and recent QSO LFs \citep{Schindler2023}, we presented the first characterization of the UV luminosity
distribution for QSOs in massive host galaxies and quantified [CII] LFs for UV-bright ($M_{UV,AB}<-26$) and UV-faint ($M_{UV,AB}>-24.5$) QSOs.  While some (3\%) massive-host QSOs are UV-bright, $\gtrsim$85\% are fainter than $-$24.5 mag. The volume density of UV-faint ($M_{UV,AB}>-24.5$) QSOs of a given host galaxy mass appears to be $\sim$35$_{-10}^{+14}$$\times$ higher  than that of UV-bright ($M_{UV,AB}<-26$) QSOs with similar-mass host galaxies.

Possible explanations for the broad UV luminosity range for QSOs with a given host mass include varying dust obscuration, a wide range of accretion efficiencies, and the range of BH masses.  While a definitive answer requires spectroscopy, we can already make a preliminary assessment of the probable drivers by extrapolating published $M_{BH}$ and $\lambda_{Edd}$ results on 34
[CII]-luminous ($L_{\text{[CII]}}>1.8\times10^9$ $L_{\odot}$) moderately UV-bright ($-28<M_{UV}<-25$) QSOs at $z>5.9$
to fainter UV luminosities.

Based on apparent trends in $M_{BH}$ and $\lambda_{Edd}$, we estimate a median $\log_{10}$$M_{BH}/M_{\odot}\sim
8.1\pm0.4$ in UV-faint QSOs in massive host galaxies (Fig.~\ref{fig:mbhdist}) while
the median $\log_{10} \lambda_{Edd}$ is $\sim-0.6\pm0.4$ (Fig.~\ref{fig:erdist}). Because $>$85\% of QSOs in massive host galaxy systems are
UV-faint, this suggests that SMBHs in UV-bright QSOs are
$\sim$15$_{-9}^{+25}\times$ higher in mass than the more typical host galaxy
system with $\log_{10}$$M_{BH}/M_{\odot}\sim8.1$ at z$\sim$6.  These results suggest that the median $M_{BH}/M_{*,galaxy}$ relation at $z\sim6$-7 is consistent with that of the local Universe \citep{Kormendy2013}.

We recognize the large uncertainties that exist in the extrapolation of $M_{BH}$ and $\lambda_{Edd}$ results to lower UV luminosities. It is therefore essential
to obtain sensitive spectroscopy for a much larger number of UV-faint QSOs at z$\geq$6 to derive
$M_{BH}$ and $\lambda_{Edd}$.  Progress is already underway with the JWST in this area
\citep[e.g.,][]{Onoue2021,Ding2023,Onoue2024} and with the Aether program \citep{Farina-aether}, but the focus on the UV-faint QSOs that are hosted by particularly massive galaxies could be stronger. From the comparisons between the two populations (UV-bright versus UV-faint), the community will not only gain a better understanding of SMBH growth in UV-bright QSOs, but should also have a comprehensive view of the full phenomenology of this growth during the earliest phases of the Universe.

In the near future, we expect to make even more progress in understanding QSO demographics by comparing the current [CII] LFs for UV-bright and UV-faint QSOs with [CII] LFs derived on the basis of z$\sim$7 galaxy samples such as for the Reionization Era Bright Emission Line Survey (REBELS; \citealt{Bouwens2022_REBELS}).  This is expected to provide us with key information on the duty cycle of QSOs in massive host galaxies, which will be essential to construct an accurate model of SMBH growth in the early Universe.

\begin{acknowledgements}
MA acknowledges support from ANID Basal Project FB210003 and ANID
MILENIO NCN2024\_112.  E.P.F. is supported by the international Gemini
Observatory, a program of NSF NOIRLab, which is managed by the
Association of Universities for Research in Astronomy (AURA) under a
cooperative agreement with the U.S. National Science Foundation, on
behalf of the Gemini partnership of Argentina, Brazil, Canada, Chile,
the Republic of Korea, and the United States of America.  This paper
is based on data obtained with the ALMA Observatory, under the program
2023.1.00443.S. ALMA is a partnership of ESO (representing its member
states), NSF(USA) and NINS (Japan), together with NRC (Canada), MOST
and ASIAA (Taiwan), and KASI (Republic of Korea), in cooperation with
the Republic of Chile. The Joint ALMA Observatory is operated by ESO,
AUI/NRAO and NAOJ.
\end{acknowledgements}

\bibliographystyle{aa} 
\bibliography{aa56183-25}

\onecolumn
\begin{appendix}

\section{Sample of [C II]-luminous QSOs with $M_{BH}$ measurements used for extrapolation to lower UV luminosities}

As part of this study, we determined that only a small fraction of
the QSOs in massive host galaxies were UV-bright and the vast
majority were UV-faint.  To make a first estimate of how the
characteristics of these fainter QSOs compare with the brighter QSOs, we
made use of a selection of QSOs which showed [CII] luminosities in
excess of 1.8$\times$10$^{9}$ $L_{\odot}$ (and thus appear to live in
a massive host galaxy) and which also have measurements of $M_{BH}$
from near-IR spectroscopy.  We present the QSOs we use for these
purposes in Table~\ref{tab:mbhsample}.  For convenience, the QSOs
presented in this table are ordered by the inferred UV
luminosity.\footnote{We do not include the lower luminosity QSOs from
\citet{Onoue2019} here due to these QSOs having [CII] luminosities
below the 1.8$\times$10$^9$ $L_{\odot}$ threshold for identifying
massive host galaxies.}

\begin{table*}[h]
\caption{$M_{BH}$ measurements available for [CII]-luminous QSOs to derive trends vs. UV luminosity.\vspace{-0.3cm}}
\label{tab:mbhsample}      
\centering          
\begin{tabular}{c c c c c c c c c c}      
  \hline\hline
         & Right     &      & $M_{UV}$ &             & $L_{\text{[CII]}}$         & FWHM$_{[\text{CII}]}$ & f$_{cont}$ & $M_{BH}$ &      \\
 QSO ID  & Ascension & Decl & (mag)    & z$_{[\text{CII}]}$ & $/10^9$$L_{\odot}$  & $/(\textrm{km/s})$ & $/$mJy & $/10^8$ $M_{\odot}$ & Ref \\
  \hline
J0100+2802   &  01:00:13.03 &    +28:02:25.8 &  $-$29.09 &  6.327 &  3.8$\pm$0.2 &  405$\pm$20 &  1.4$\pm$0.1 &  97.34 & [5]\\
J2310+1855   &  23:10:38.88 &    +18:55:19.7 &  $-$27.80 &  6.003 &  8.7$\pm$1.4 &  393$\pm$21 &  8.3$\pm$0.6 & 21.91 & [5]\\
J025$-$33    &  01:42:43.72 &  $-$33:27:45.6 &  $-$27.76 &  6.337 &  5.7$\pm$0.2 &  370$\pm$16 &  2.5$\pm$0.1 &  25.89 & [1]\\
J1148+5251   &  11:48:16.65 &    +52:51:50.4 &  $-$27.56 &  6.419 &  4.2$\pm$0.4 &  287$\pm$28 &  1.8 &  52.99 &  [2]\\
J0706+2921   &  07:06:26.38 &    +29:21:05.5 &  $-$27.44 &  6.604 &  2.2$\pm$0.3 &  413$\pm$36 &  0.7$\pm$0.1 &  21.13 & [4]\\
PSOJ158$-$14 &  10:34:46.50 &  $-$14:25:15.9 &  $-$27.41 &  6.069 & 11.5         &  780$\pm$27 &  3.5$\pm$0.1 &  17.04 & [5]\\
P036+03      &  02:26:01.87 &    +03:02:59.2 &  $-$27.28 &  6.541 &  3.4$\pm$0.1 &  237$\pm$7  &  2.6$\pm$0.1 &  37.37 & [5]\\
J1509$-$1749 &  15:09:41.78 &  $-$17:49:26.8 &  $-$27.14 &  6.123 &  2.3         &  631$\pm$70 &  1.7$\pm$0.1 &  26.44 & [5]\\
P231$-$20    &  15:26:37.84 &  $-$20:50:00.9 &  $-$27.14 &  6.587 &  3.5$\pm$0.3 &  393$\pm$35 &  4.4$\pm$0.2 &  40.75 & [5]\\
J0038$-$1527 &  00:38:36.10 &  $-$15:27:23.6 &  $-$27.13 &  7.034 &  3.2$\pm$0.3 &  339$\pm$25 &  1.0$\pm$0.1 &  13.56 & [4]\\
J1319+0950   &  13:19:11.29 &    +09:50:51.5 &  $-$26.99 &  6.135 &  4.0$\pm$0.4 &  532$\pm$57 &  5.1$\pm$0.2 &  18.84 & [5]\\
P183+05      &  12:12:26.97 &    +05:05:33.5 &  $-$26.99 &  6.439 &  7.2$\pm$0.3 &  397$\pm$19 &  4.8$\pm$0.2 &  30.08 & [5]\\
J2002$-$3013 &  20:02:41.59 &  $-$30:13:21.7 &  $-$26.90 &  6.688 &  2.2$\pm$0.2 &  308$\pm$36 &  3.2$\pm$0.2 &  16.22 & [4]\\
PSOJ011+09   &  00:45:33.57 &    +09:01:56.9 &  $-$26.85 &  6.470 &  2.2         &  449$\pm$66 &  1.2$\pm$0.1 &  6.3 & [4]\\
P359$-$06    &  23:56:32.45 &  $-$06:22:59.3 &  $-$26.74 &  6.172 &  2.6$\pm$0.1 &  341$\pm$18 &  0.8$\pm$0.1 &  11.23 & [5]\\
J1007+2115   &  10:07:58.26 &    +21:15:29.2 &  $-$26.73 &  7.515 &  2.8$\pm$0.2 &  349$\pm$56 &  3.0$\pm$0.1 & 14.32 & [4]\\
J0923+0402   &  09:23:47.12 &    +04:02:54.6 &  $-$26.68 &  6.633 &  2.0$\pm$0.2 &  350$\pm$36 &  0.3$\pm$0.1 &  17.74 & [4]\\
J0224$-$4711 &  02:24:26.54 &  $-$47:11:29.4 &  $-$26.67 &  6.522 &  5.4$\pm$0.2 &  334$\pm$15 &  2.4$\pm$0.1 &  21.88 & [3]\\
J0252$-$0503 &  02:52:16.64 &  $-$05:03:31.8 &  $-$26.63 &  7.001 &  2.7$\pm$0.5 &  393$\pm$96 &  1.2$\pm$0.1 &  12.84 & [4]\\
J0910$-$0414 &  09:10:54.54 &  $-$04:14:06.9 &  $-$26.61 &  6.636 &  4.2$\pm$0.2 &  783$\pm$40 &  3.7$\pm$0.1 &  36 & [4]\\
J0411$-$0907 &  04:11:28.63 &  $-$09:07:49.7 &  $-$26.58 &  6.826 &  2.1$\pm$0.5 &  371$\pm$67 &  0.3$\pm$0.1 &  9.48 & [4]\\
J2318$-$3029 &  23:18:33.10 &  $-$30:29:33.6 &  $-$26.39 &  6.146 &  2.2$\pm$0.1 &  293$\pm$17 &  3.1$\pm$0.1 &  14.56 & [5]\\
P308$-$21    &  20:32:10.00 &  $-$21:14:02.4 &  $-$26.29 &  6.236 &  3.4$\pm$0.2 &  541$\pm$32 &  1.2$\pm$0.1 &  16.79 & [5]\\
J1135+5011   &  11:35:08.92 &    +50:11:32.6 &  $-$26.16 &  6.585 &  1.9$\pm$0.1 &  425$\pm$75 &  0.2$\pm$0.1 &  14.85 & [4]\\
J2054$-$0005 &  20:54:06.50 &  $-$00:05:14.6 &  $-$26.15 &  6.039 &  3.1$\pm$0.1 &  236$\pm$12 &  3.2$\pm$0.1 &  14.81 & [5]\\
P338+29      &  22:32:55.15 &    +29:30:32.2 &  $-$26.14 &  6.658 &  2.0$\pm$0.1 &  740$\pm$310&  1.0$\pm$0.2 &  30.58 & [4]\\
J0305$-$3150 &  03:05:16.92 &  $-$31:50:55.8 &  $-$26.13 &  6.614 &  5.9$\pm$0.4 &  225$\pm$15 &  5.3$\pm$0.2 &  5.4 & [5]\\
J1048$-$0109 &  10:48:19.08 &  $-$01:09:40.4 &  $-$25.96 &  6.676 &  2.1$\pm$0.2 &  299$\pm$24 &  2.8$\pm$0.1 &  22.94 & [5]\\
J1058+2930   &  10:58:07.72 &    +29:30:41.7 &  $-$25.68 &  6.585 &  2.7$\pm$0.5 &  336$\pm$46 &  0.4$\pm$0.1 & 5.42 & [4]\\
J0109$-$3047 &  01:09:53.14 &  $-$30:47:26.3 &  $-$25.59 &  6.790 &  1.9$\pm$0.2 &  354$\pm$34 &  0.5$\pm$0.1 &  11.02 &  [5]\\
J0525$-$2406 &  05:25:59.68 &  $-$24:06:23.0 &  $-$25.47 &  6.540 &  5.5$\pm$0.2 &  259$\pm$16 &  4.3$\pm$0.1 & 2.93 & [4]\\
J0246$-$5219 &  02:46:55.90 &  $-$52:19:49.9 &  $-$25.36 &  6.888 &  3.1$\pm$0.2 &  400$\pm$29 &  2.6$\pm$0.1 & 10.51 & [4]\\
J0319$-$1008 &  03:19:41.66 &  $-$10:08:46.0 &  $-$25.36 &  6.828 &  2.5$\pm$0.7 &  727$\pm$205 & 0.1$\pm$0.1 & 3.98 & [4]\\
J0910+1656   &  09:10:13.65 &    +16:56:30.2 &  $-$25.34 &  6.729 &  1.9$\pm$0.3 &  379$\pm$50 &  0.2$\pm$0.1 & 4.09 & [4]\\
J1243+0100   &  12:43:53.93 &    +01:00:38.5 &  $-$24.13 &  7.070 &  2.5$\pm$0.1 &  280$\pm$12 &  1.5$\pm$0.1 & 3.3  & [6]\\
\hline
\end{tabular}
\tablefoottext{a}{References: [1] = \citet{Chehade2018}, [2] = \citet{Shen2019}, [3] = \citet{Reed2019}, [4] = \citet{Yang2021}, [5] = \citet{Farina2022}, [6] = \citet{Matsuoka2019_faintz7qso}}
\end{table*}

\end{appendix}
\end{document}